# Electrical capacitance volume sensor for microgravity mass gauging: advancements in sensor calibration for microgravity fluid configurations and propellant management

Matthew A. Charleston[1*], Shah M. Chowdhury[1], Benjamin J. Straiton[1], Qussai M. Marashdeh[1], and Fernando L. Teixeira[2]

[1] Tech4Imaging, LLC, Columbus, OH, USA
[2] ElectroScience Laboratory, The Ohio State University, Columbus, OH, USA

[*]E-mail: m.charleston@tech4imaging.com

**Abstract.** Microgravity mass gauging has gained increasing importance in recent years due to the acceleration in planning for long term space missions as well as in-space refueling & transfer operations. It is of particular importance with cryogenic propellants where periodic tank venting maneuvers and leak detection place a special emphasis on accurate mass gauging. Several competing technologies have arisen, but capacitance mass gauging has several distinct advantages due to its low mass, non-intrusiveness, and whole volume interrogation technique. Capacitance based measurement has also seen recent success in measuring cryogenic liquid nitrogen and hydrogen volume fraction and flow rate. However, the effects of gravity on fluid behavior make the calibration and testing of these sensors difficult on the ground. In this paper a prototype sensor is constructed that can emulate fluid positions in microgravity and gravity configurations. Experimental propellant fills and drains are conducted using a simulant fluid with similar electrical properties to cryogenic propellants. This expanded dataset is compared with previous simulation results and used to construct a machine learning model capable of calculating the fluid mass in tanks both with and without propellant management devices.

## 1. Background and Introduction

Microgravity mass gauging is an active area of research at NASA to facilitate the use of cryogenic propellants in long term missions involving in-space refueling or in-situ resource utilization [1]. Historically, methods involving level sensing techniques and propellant mass flow integration (bookkeeping) were used to estimate propellant mass [2]. However, there are major drawbacks to these techniques. Bookkeeping techniques lose accuracy after repeated thruster burns, increasing uncertainty as the need for accurate mass gauging becomes most critical and level gauging requires settled propellant configurations. These techniques cannot be used for the more complex future missions involving cryogenic propellant tank-to-tank transfer, fluid sloshing, microgravity surface tension dominated fluid configurations, and fluid boiling.

Current frontrunners for microgravity mass gauging include Electrical Capacitance Volume Sensing (ECVS) [3] [4], Radio Frequency Mass Gauging (RFMG) [5], and Acoustic Modal Gauging

(AMG) [6], among others [2]. RFMG and AMG excite the tank using radio frequency and acoustic waves respectively and measure the spectral response. The response modes are affected by the fluid, and this is used to estimate the propellant mass fraction. The major drawback of these techniques is that they are model-based [7]. The relationship between measured modes and propellant mass is dependent on the propellant position in addition to the propellant mass. The mass estimation is based on simulation or experimental data of the microgravity settled fluid distribution in the tank. Under this assumption differences between the simulated or experimental calibration data and the deployed system will result in erroneous readings. Different heat loads, dynamic fluid motion, and metastable fluid configurations all may affect the accuracy of these techniques [7]. In order to cope with these various aspects of future space missions, a mass gauging method that is not model-based is desired.

ECVS is a capacitance-based sensing technique that uses the 3D volumetric sensitivity of electric fields to measure the mass fraction of fluids inside a tank. Electrode plates are arrayed on the periphery of a tank and the inter-electrode capacitances are measured. In previous studies the authors prototyped and optimized the sensor design [8], and analyzed different mass fraction estimation algorithms in simulation [3]. Ultimately, a 12-electrode dodecahedron design using machine learning algorithms for mass fraction reconstruction was determined to perform well. ECVS sensors using 12 plates have 66 independent measurements. This data includes information on the fluid position and mass, allowing the mass measurement to be compensated for changes in fluid position. In this study, experimental data is taken using a prototype dodecahedron sensor and used to develop a robust calibration that is capable of handling various fluid positions and responding in real-time.

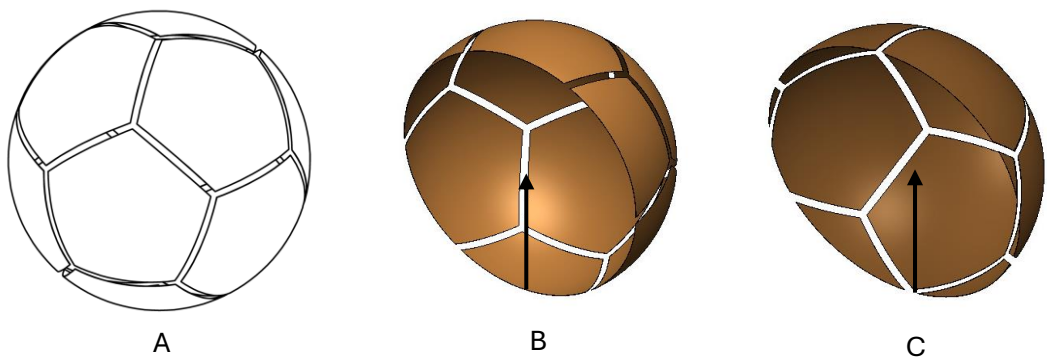

**Figure 1.** A) Dodecahedron electrode plate layout. Twelve plates are arrayed around the tank as a spherical projection of the sides of a dodecahedron. B) Plate-center fill orientation. C) Gap-center fill orientation

## 2. Experimental Setup

*2.1 Sensor Design and Construction*

The sensor electrode layout is displayed in Figure 1. This design was previously tested against an octahedron style design and the increased rotational symmetry of the design provided a more stable signal as fluid changed position inside the tank allowing for more accurate mass gauging.

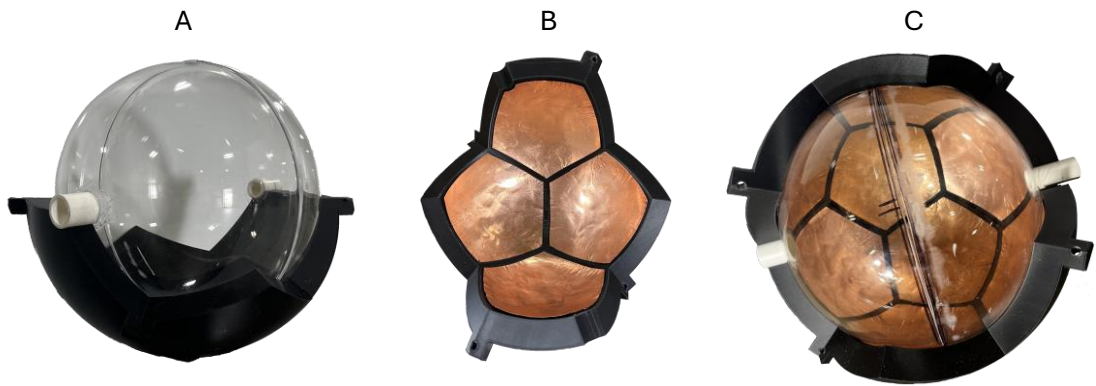

**Figure 2.** A) Acrylic tank with one 3D printed segment. B) One segment shown alone with copper electrodes attached. C) Two segments with copper electrodes mounted to tank

The sensor was constructed to clamp onto a 9.8" outer diameter, ¼" thick spherical acrylic tank. The tank, electrodes, and clamp-on segments are shown in figure 2. The tank was constructed with ports at the top and bottom for filling and draining the fluid. The 3 sensor segments clamp around the tank and drain ports such that the electrodes are pressed against the outer diameter of the tank. The electrode segments are 3D printed to be 1" thick and a ground screen is applied on the outer layer to prevent interference in the capacitance signal.

## 2.2 Experimental Test Cases

The sensor is calibrated by conducting several tests where the sensor is filled and drained in different orientations. Mineral oil is used as a simulant fluid as it has a dielectric constant of 2.16, similar to the dielectric constants of cryogenic propellants 1.2-1.6. During each test a scale is used to record the mass of fluid in the tank, so the capacitance measurements can later be associated with the mass fraction through a machine learning algorithm. First, stratified fill tests are conducted. The sensor is filled up in discrete increments and gravity causes the fluid to stay in a stratified level. Two experiments are conducted as shown in figure 1, one where the bottom of the sensor is the gap between 3 electrode plates, and the other where the bottom of the sensor is the center of an electrode plate. These are referred to as stratified gap-center and stratified plate-center respectively.

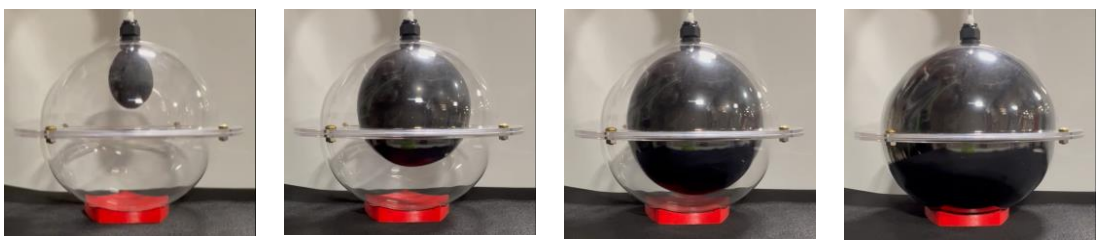

**Figure 4.** Balloon fill illustrated during different stages of inflation. Balloon pushes fluid to the outside walls of the tank, similar to how surface tension arranges fluid in microgravity

Next, a settled microgravity fluid fill is approximated using a balloon. Shown in figure 4, a latex balloon is inserted into the top of the sensor while the sensor is filled with oil. The balloon is periodically inflated causing an air bubble to appear towards the center of the sensor and the fluid is distributed toward the outer walls. As the balloon inflates the mass of the fluid ejected through the drain port is measured for reference.

Finally, dynamic fluid cases are recreated through sloshing experiments shown in figure 5. The tank and data acquisition system are mounted on a cart and sloshed at different fill levels. Different slosh intensities were conducted ranging from gentle oscillations of the fluid level to violent splashing so that a wide range of fluid distributions could be recorded for a given mass fraction of fluid.

*2.3 Simulation Test Cases*

To conduct testing with fluid distributions that could not easily be created in the laboratory, fluid distributions were simulated, and sensor readings recorded in COMSOL Multiphysics. Simulations were conducted to repeat the stratified gap-center and stratified plate-center test cases. Additionally, annular cases where a spherical vapor bubble is located in the center of the tank and core annular cases where a spherical propellant bubble is located in the center of the tank. Offsets configurations of the annular and core annular cases were also simulated. Additional detail can be found in a previous work where the simulations were initially conducted [3]. In total, 11, 280 fluid configurations were simulated.

## 3. Machine Learning Algorithm

*3.1 Data Conditioning*

The collected experimental data was conditioned so it could be fed in as training data to a machine learning algorithm. The balloon test data was collected in real-time because reactions between the mineral oil and the latex balloons caused very short lifespans for the balloons once immersed. Capacitance data was collected continuously and synchronized with the scale. Each time the balloon was inflated the system was allowed to equilibrate for around 30 seconds before continuing on. The equilibrated test points were then filtered out from the dynamic data. The ballon, stratified gap-center, and stratified plate-center fill tests were then increased in density using a MATLAB PCHIP interpolation. Originally datapoints were taken in intervals of approximately 3%, but after interpolation datapoints were available every 0.1%. The data from sloshing experiments was not conditioned in any way.

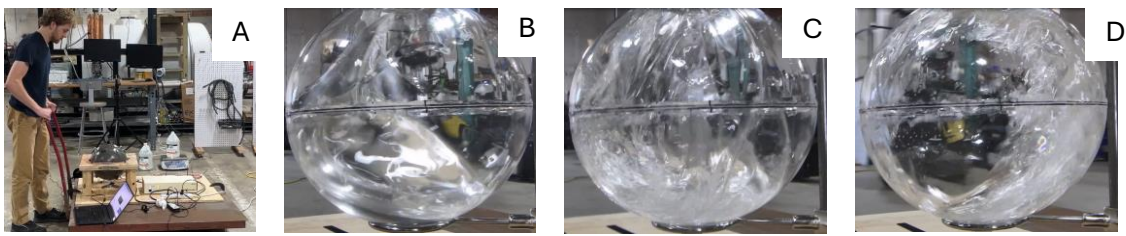

**Figure 5.** A) Sloshing cart, research technician, sensor, and data acquisition system. B-D) Various fluid states achieved during a slosh test illustrated for reference. In all images B-D, the tank is filled with the same mass fraction of mineral oil, approximately 25%.

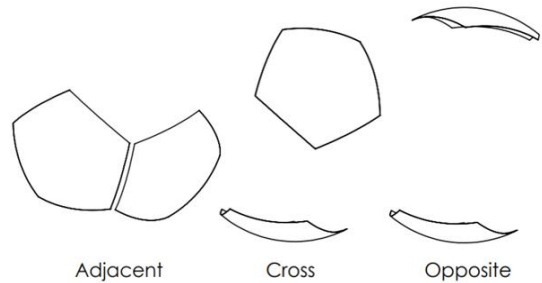

**Figure 6.** Channel Types of Dodecahedron Sensor

*3.1 Algorithm*

Due to the rotational symmetry of the dodecahedron design, there are only 3 fundamentally different orientations of electrode pairs, adjacent, cross, and opposite, illustrated in figure 6. These configurations are referred to as the channel type. Of the 66 independent measurements in the dodecahedron design, 30 are adjacent, 30 are cross, and 6 are opposite channels. When all 66 measurements are considered individually, the sensor has sufficient data to determine the 3D position of objects through a technique known as electrical capacitance volume tomography.

However, the sensitivity of the capacitance measurement depends mainly on the radial distribution of the fluid, as adjacent channels have a higher sensitivity in a more localized area than cross or opposite channels. By averaging the 66 measurements into 3 groups based on channel type, this radial distribution data is preserved, but the rotational distribution data is lost. This is beneficial for calibration of the sensor, as fluid fill types do not have to be repeated for every rotational orientation of the device.

All datasets discussed previously are averaged by channel type and used as training data in a Bagged Decision Tree using MATLAB's Statistics and Machine Learning Toolbox and 5-fold cross validation. In total 223,649 input groups were considered and an RMSE of 0.0029 was achieved.

**4. Results and Discussion**

Validation of the dataset on the 3 experimental fill tests showed errors withing ±3.61% and validation on the slosh datasets showed similar errors within ±3.64%. Additional validation was conducted on data outside the training set. Stationary data points collected in-between slosh tests showed error within ±0.71%. Finally, real time data from the balloon test was analyzed in figure 7. The selected dataset included a balloon popping event around time 13:58.5, where the fluid orientation rapidly changed from an annular to stratified configuration. The overall error from this dataset was within ±4.75%. If the outliers are removed outside of 3 standard deviations, the error metric becomes ±3.79%. This dataset has some additional errors caused by delays between the balloon expansion and the fluid traveling out of the sensor, through a short tube, and onto the scale.

To test if the algorithm works with propellant management devices, acrylic vanes were simulated in COMSOL. The machine learning algorithm was used on this data and the error was within ±3.10%. Using composite tanks and composite vanes, this technique demonstrates excellent results. Investigations with metallic vanes and metallic tanks are ongoing.

Overall, the accuracy of all tests is around ±4%. The errors metrics in this investigation are much more conservative than previous investigations of microgravity mass gauging techniques, as dynamic cases are considered. Analysis of the differences between experimental and simulation data revealed differences in the range of 3-4% and the repeatability error between tests was also around 3-4%. Based on this information, the accuracy seems to be limited by the accuracy of the simulation, the current data acquisition system, and experimental techniques. Improvements in these areas would likely result in improvements in mass gauging accuracy.

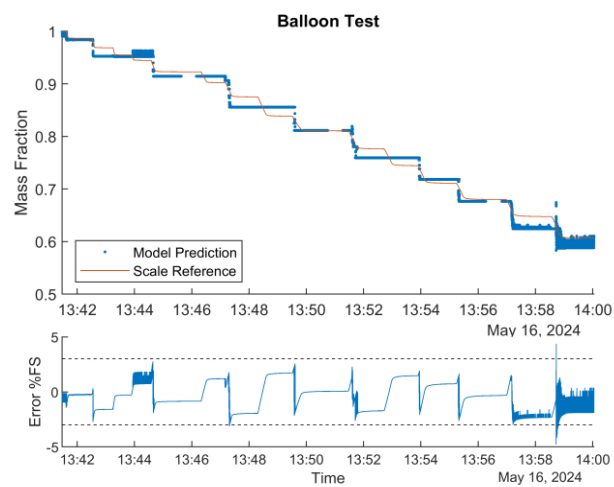

**Figure 7.** Balloon test data. Machine learning model prediction is compared to the reference scale and the live error is computed. Dashed lines in the error plot are drawn at ±3%

## References

[1] National Aeronautics and Space Administration, "2020 NASA Technology Taxonomy," National Aeronautics and Space Administration, 2020.

[2] B. Yendler, "Review of Propellant Gauging Methods," in *44th AIAA Aerospace Sciences Meeting and Exhibit*, Reno, Nevada, 2006.

[3] S. M. Chowdhury, M. A. Charleston, Q. M. Marashdeh and F. L. Teixeira, "Propellant Mass Gauging in a Spherical Tank under Micro-Gravity Conditions Using Capacitance Plate Arrays and Machine Learning," *Sensors,* vol. 23, no. 20, p. 8516, 2023.

[4] J. Storey, B. Marsell, M. Elmore and S. Clark, "Preliminary Results from Propellant Mass Gauging in Microgravity with Electrical Capacitance Tomography," NASA, 2022.

[5] G. A. Zimmerli, K. R. Vaden, M. D. Herlacher, D. A. Buchanan and N. T. Van Dresar, "Radio Frequency Mass Gauging of Propellants," NASA/TM-2007-214907, Cleveland, 2007.

[6] K. M. Crosby, R. J. Werlink and E. A. Hurlbert, "Liquid Propellant Mass Measurement in Microgravity," *Gravitational and Space Research,* vol. 9, no. 1, pp. 50-61, 2021.

[7] J. Parson, K. Roe, J. Feller, M. Khasin and S. P. Sharma, "Mass Gauging for Sustained Presence in Outer Space: A Technology Review," in *AIAA Scitech 2021 Forum*, 2021.

[8] M. A. Charleston, S. M. Chowdhury, Q. M. Marashdeh, B. J. Straiton and F. L. Teixeira, *Towards Microgravity Mass Gauging in Spherical Tanks using ECVT,* Kailua-Kona, HI: 30th Space Cryogenics Workshop, 2023.